\documentclass[aps,PRL,preprint,superscriptaddress]{revtex4}

\usepackage[utf8]{inputenc}

\usepackage{natbib}
\usepackage{graphicx}
\usepackage{dcolumn}
\usepackage{bm}
\usepackage{epsfig}
\usepackage{textcomp}
\usepackage{gensymb}
\usepackage[english,iso]{isodate}
\usepackage[colorlinks]{hyperref}
\usepackage{etoolbox}

\begin{document}

\title{The true corrugation of a h-BN nanomesh layer}

\author{L. H. de Lima}
\affiliation{Paul Scherrer Institut, 5232 Villigen PSI, Switzerland}
\affiliation{Centro de Ci{\^e}ncias Naturais e Humanas, Universidade Federal do ABC, Santo André, 09210-580, Brazil}
\author{T. Greber}
\affiliation{Physik-Institut, Universit{\"a}t Z{\"u}rich, 8057 Z{\"u}rich, Switzerland}
\author{M. Muntwiler}
\affiliation{Photon Science Division, Paul Scherrer Institut, 5232 Villigen PSI, Switzerland}

\date{\today, co-author review}

\begin{abstract}
Hexagonal boron nitride (h-BN) nanomesh, a two-dimensional insulating monolayer, grown on the (111) surface of rhodium exhibits an intriguing hexagonal corrugation pattern with a lattice constant of 3.2 nm.
Despite numerous experimental and theoretical studies no quantitative agreement has been found on some details of the adsorption geometry such as the corrugation amplitude.
The issue highlights the differences in chemical and electronic environment in the strongly bound pore regions and the weakly bound wire regions of the corrugated structure.
For reliable results it is important to probe the structure with a method 
that is intrinsically sensitive to the position of the atomic cores rather than the electron density of states.
In this work, we determine the corrugation of h-BN nanomesh from angle- and energy-resolved photoelectron diffraction measurements with chemical state resolution.
By combining the results from angle and energy scans
and comparing them to multiple-scattering simulations
true adsorbate-substrate distance can be measured with high precision, 
avoiding pitfalls of apparent topography observed in scanning probe techniques.
Our experimental results give accurate values for the peak to peak corrugation amplitude (0.80\,\AA), 
the bonding distance to the substrate (2.20\,\AA) and 
the buckling of the boron and nitrogen atoms in the strongly bound pore regions (0.07\,\AA).
The results are important for the development of theoretical methods involving the quantitative description of van der Waals systems like it requires the understanding of the physics of two-dimensional sp$^2$ layers.
\end{abstract}

\maketitle

\begin{figure}[ht]
	\centering
	\includegraphics[width=8.8cm]{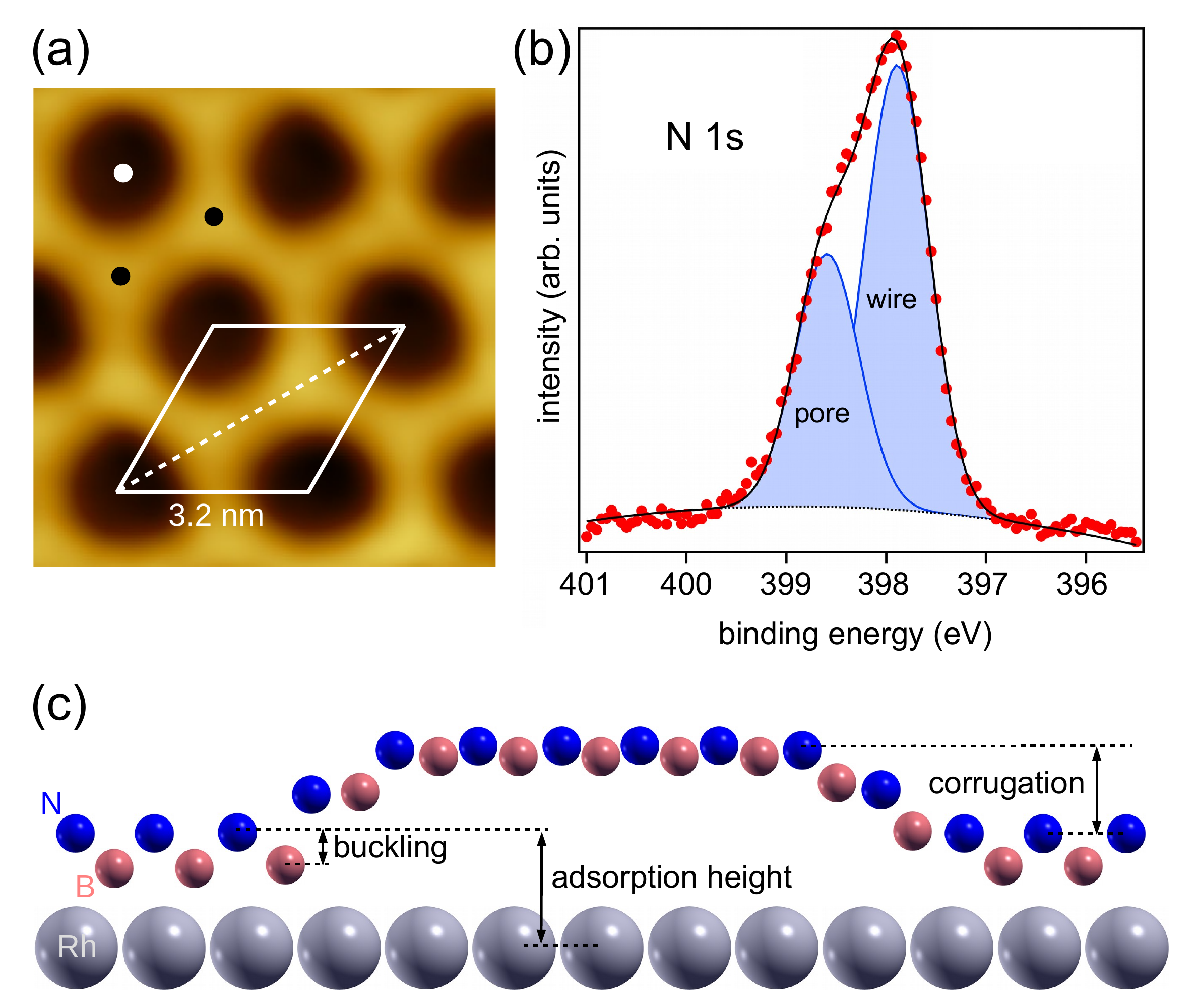}
	\caption{\label{fig1}%
		(a) STM topography image of h-BN nanomesh on Rh(111), $I_t = 10 \, \mathrm{pA}$, $V_t = 1.0 \, \mathrm{V}$.
		The dark regions are depressions (``pores'') in the continuous h-BN layer, 
		the bright regions are the elevated ``wires''. 
		The superstructure unit cell is highlighted, the lattice parameter is 3.2 nm. 
		The white and black dots mark the high-symmetry points in the pore and on the wires.
		(b) N 1s core-level XPS spectrum measured at $h\nu = 565 \, \mathrm{eV}$ and normal emission. 
		The plot shows experimental data (red dots) and a curve-fit with a two-component Gaussian profile
		(lines and shaded areas).
		c) Schematic vertical cut through the layer along the main diagonal of the supercell (dashed line in (a)). 
		The N, B and Rh atoms are represented by blue, peach and gray spheres, respectively. 
		Note the different registry in the pore and wire regions (drawing not to scale) of the 13-on-12 superstructure.
	}
\end{figure}

Two-dimensional materials like graphene \cite{Novoselov2004}, phosphorene \cite{LiL14NN}, hexagonal boron nitride (h-BN) \cite{Nagashima1995} and transition metal dichalcogenides (TMD) \cite{Manzeli2017}
host a wealth of fascinating electronic properties, 
such as a relativistic band dispersion with Dirac fermions, 
topologically non-trivial band structure, spin quantum Hall effect, 
or unconventional superconductivity \cite{CaoY18N}. 
In technology, heterostructures of 2D van der Waals materials are promising routes to novel device architectures, 
on the one hand due to their well-defined atomic layers and sharp interface, 
and on the other hand due to the possibility of tuning and exploiting the mentioned quantum effects \cite{Wang2011,Xu2013,IannacconeG18NN,LiuY19N}.
Due to its non-interacting, insulating, and structural properties, 
h-BN serves as an ideal interface to graphene \cite{Dean2010}, 
topological insulators \cite{Gehring2012}, 
as well as dichalcogenides \cite{Zhang2015,Yan2015,Zhang2016,Chen2018}.
Moreover, when a monolayer of h-BN is grown on the surface of certain transition metals, 
it may form a corrugated superstructure with an in-plane lattice constant of the order of nanometers \cite{AuwarterW18SSR,Preobrajenski2007,Brugger2007,Preobrajenski2008}.
The corrugated superstructure arises due to a competition 
between lattice matching with the substrate and strong bonding within the sp$^2$ layer. 
Most prominent examples are h-BN/Rh(111) \cite{Corso2004,Muller2009,Martoccia2010_Rh111,Roth2011} and h-BN/Ru(0001) \cite{Paffett1990,Goriachko2007,Martoccia2010_Ru0001}.
Though the layer forms a continuous sheet, 
the corrugated structure is usually called a {\it nanomesh} 
due to its porous appearance in STM (Fig.~\ref{fig1}(a)).
The distinct electronic properties in the {\it pore} and {\it wire} regions modulate the local work function,
which allows the corrugated layer to serve as a nanotemplate 
for the self-assembly of atoms, molecules and clusters 
\cite{Berner2007,Dil2008,Brihuega2008,Ma2010,Ding2011,Patterson2014,Iannuzzi2014}.
Despite the recent scientific and technological advances,
the atomic and electronic details of the interface geometry such as layer distance 
and the structural or electronic nature of the observed corrugation are not well understood.
For instance, in h-BN/Cu(111) an earlier study explains the moir{\'e} pattern seen in scanning tunnelling microscopy 
as a purely electronic effect, i.e., a modulation of the local work function \cite{Joshi12NL},
whereas two independent x-ray standing wave studies report atomic corrugation \cite{SchwarzM17AN,Brulke2017}
(see also the discussion in \cite{AuwarterW18SSR}).

In this article, we study the adsorption height and corrugation of h-BN nanomesh on Rh(111) using photoelectron diffraction (XPD/PhD).
Photoelectron diffraction is inherently sensitive to the distance between atomic cores and less sensitive to the valence structure.
The combination of modern, synchrotron-based XPD with high-performance multiple scattering calculations
allows to exploit the backscattering regime
and clarify structural details where other methods remain insensitive or inaccurate.

For h-BN on Rh(111), the superstructure consists of $13\times13$ unit cells of h-BN on top of $12\times12$ unit cells of Rh(111) with an in-plane superlattice parameter of 3.2 nm \cite{Corso2004,Bunk2007,Laskowski2007}, cf.~Fig.~\ref{fig1}. 
Within the unit cell a depression forms
where the interaction between the h-BN and Rh is strong ({\it pores})
and is surrounded by a network of suspended regions ({\it wires}), 
where the interaction is weak \cite{Laskowski2007,Berner2007}.
Despite the extensive list of publications on h-BN/Rh(111) reported in the literature (cf.\ Table \ref{tab1}),
the vertical positions of the B and N atoms as well as the lateral size of the pores are still a matter of controversy.
Various levels of density functional theory (DFT) fail to give satisfactory predictions,
as the result depends strongly on the choice of exchange-correlation functional.

On the experimental side, few techniques are capable of producing absolute measurements of atomic coordinates of the surface layers.
Widely used scanning probe techniques (STM, AFM) measure the corrugation of a surface by following the contour of a specific response with a mechanically adjusted metal tip.
However, since the tip is mainly sensitive to the electronic local density of states (LDOS)
and since the electronic structure of the nanomesh in the pores and wires is different, 
discrepancies in the probed heights can be expected \cite{McKee2015}.
In the case of STM, the observed corrugation amplitude is typically of the order of \mbox{1.0 \AA}, 
but this value strongly depends on the applied tunnelling potential \cite{Brihuega2008}, 
and even contrast reversal can be observed \cite{Berner2007,Brihuega2008,KochS12JPCM}. 
Moreover, for sp$^2$ corrugated monolayers, 
artefacts in the measured atomic topography may be induced by the directionality of the $\pi$-orbitals.
These distortions vary across the unit mesh due to the curvature of the monolayer \cite{Dubout2016}. 
Finally, the adsorption height is fundamentally inaccessible with scanning probe techniques.

In the case of three-dimensional crystals, highest structural resolution is typically obtained from x-ray diffraction methods.
Two variants, surface x-ray diffraction (SXRD) and x-ray standing wave spectroscopy (XSW), 
have been applied to h-BN systems.
While the lateral superstructure of h-BN/Rh(111) is clearly resolved in SXRD,
the vertical corrugation has not been resolved 
due to the dominant signal from the high-$Z$ substrate atoms \cite{Bunk2007}.
XSW measurements have shown to give accurate layer positions of h-BN and graphene systems \cite{Hagen2016,OhtomoM17Ns,SchwarzM17AN,Brulke2017}.
The method, requiring good quality substrates as well as tunable tender x-rays (2-5 keV) at high energy resolution and high photon flux, is, however, instrumentally demanding.

\begin{table}
	\caption{\label{tab1}%
		Summary of results for the adsorption height and corrugation amplitude of h-BN/Rh(111) 
		obtained with different theoretical and experimental methods.
		The adsorption height refers to the shortest vertical distance between a nitrogen and a Rh atom.}
	\centering
	\begin{tabular}{lccc}
		\hline
		\hline
		Method  & Adsorption & Corrugation & Ref. \\
		& Height (\AA) & (\AA) & \\
		\hline
		DFT WC-GGA & 2.17 & 0.55 & \cite{Laskowski2007} \\
		DFT LDA & & 2.65 & \cite{Patterson2014} \\
		DFT optB86b vdW-DF & & 2.38 & \cite{Patterson2014} \\
		DFT GGA-PBE  & & 2.20 & \cite{Patterson2014} \\
		DFT PBE-rVV10  & 2.2 & 2.28 & \cite{Iannuzzi2014}\\
		DFT revPBE-D3  & 2.2 & 1.06 & \cite{Iannuzzi2014} \\
		DFT vdW-DF  & 3.3 & 0.90 & \cite{Iannuzzi2014} \\
		\hline
		STM & & 0.5 & \cite{Corso2004} \\
		STM C$_{60}$ decoration & & 2.0 & \cite{Corso2004} \\
		STM & & 0.7 & \cite{Goriachko2007} \\
		STM & & 0.5 & \cite{Patterson2014} \\
		STM & & 0.9 & \cite{Iannuzzi2014} \\
		nc-AFM & & 0.9 & \cite{KochS12JPCM} \\
		\hline
		XPD & 2.20 & 0.80 & this \\
		\hline
		\hline
	\end{tabular}
\end{table}

To overcome the presented difficulties, 
we investigate the system using x-ray photoelectron diffraction (XPD/PhD/PED)\cite{Fadley1987,Woodruff2007}. 
A photoelectron is emitted from a defined chemical environment and scatters at the electrostatic potential of the cores of neighbouring atoms. 
Interference of the scattered and direct waves leads to characteristic diffraction patterns as a function of electron momentum.
Experimentally, the electron momentum is scanned by variation of the emission angle (by rotation of the sample or detector in two dimensions) and/or the kinetic energy (by tuning the photon energy).
In this paper, we refer to angle-scanned measurements as XPD and to energy-scanned measurements as PhD, as usually adopted in the literature \cite{Woodruff2007},
and use the acronym PED for the more general physical concept.
PED is sensitive to the positions of the atomic cores in the top few layers of a surface.
It does not require the same long-range periodicity as x-ray diffraction or low energy electron diffraction (LEED). 
Furthermore, it is selective of the probed local environment by the binding energy of the initial state.
In the case of the commensurate $1 \times 1$ h-BN monolayer on Ni(111), 
NI-XSW \cite{OhtomoM17Ns} and PhD \cite{MuntwilerM17JSR} give almost perfect agreement.
For tuning the scattering conditions, the use of synchrotron light is necessary. 
All experiments presented in this work were carried out at the {\it Photo-Emission and Atomic Resolution Laboratory} (PEARL) beam line of the Swiss Light Source (SLS) \cite{MuntwilerM17JSR}.

\begin{figure*}[ht]
	\centering
	\includegraphics[width=\textwidth]{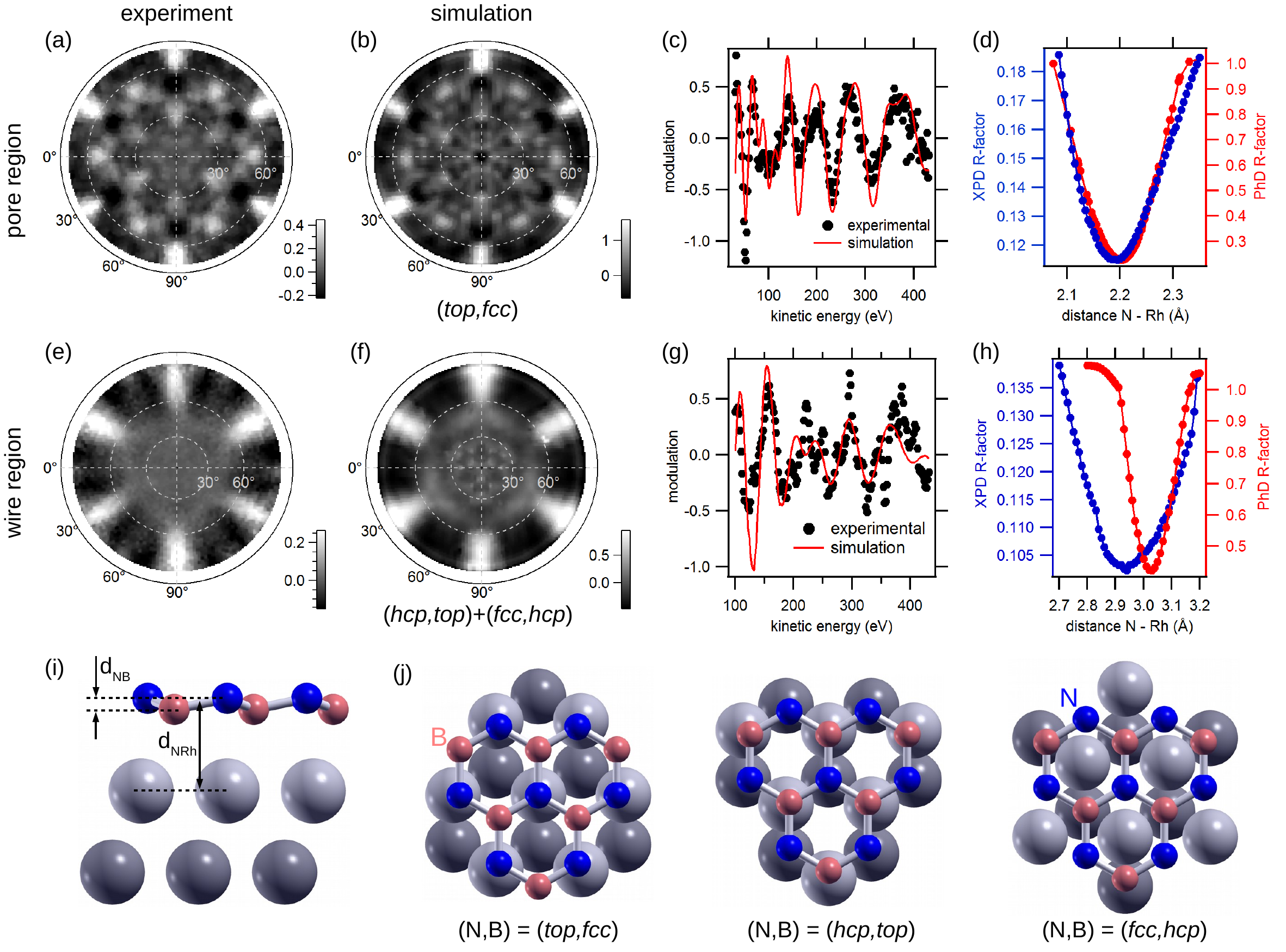}
	\caption{\label{figxpd}%
		Experimental and simulated photoelectron diffraction patterns
		of the pore component (a-d) and the wire component (e-h) of the N 1s spectrum.
		(a,e) Experimental XPD patterns (modulation of intensity versus emission angle in stereographic projection). 
		(b,f) Simulated XPD patterns for the best fitting atomic structure models. 
		(c,g) Experimental and simulated PhD modulation curves at normal emission. 
		(d,h) R-factor as a function of the distance $d_{\mathrm{NRh}}$ for both XPD (blue, left axis) and for PhD (red, right axis).
		(i) Side view of the h-BN/Rh interface.
		The adsorption height $d_{\mathrm{NRh}}$ and buckling parameter $d_{\mathrm{NB}}$ are labelled.
		(j) Top view of the three high-symmetry adsorption configurations.%
	}
\end{figure*}

A monolayer of h-BN is obtained by chemical vapour deposition (CVD), exposing the hot Rh(111) single crystal surface (1050 K) to $4.5\times10^{-7}$ mbar borazine (HBNH)$_3$ for \mbox{4 minutes} \cite{Corso2004}. 
The sample shows the well-known $12 \times 12$ corrugated morphology in STM and LEED, 
and two chemically shifted components in the N 1s photoelectron spectrum, Fig.~\ref{fig1}.
The shift of \mbox{0.7 eV} and the 2:1 intensity ratio is in accordance to earlier studies
that attribute the higher binding energy peak to the more strongly bound N atoms in the pore region \cite{Preobrajenski2007,Laskowski2007,HemmiA19NL}.
The appearance of two separable peaks in the XPS spectrum allows us to track the diffraction patterns of the two peaks, i.e., the pore and wire region separately by fitting each measured spectrum with two Gaussian profiles.
We acquire a $2\pi$ steradian XPD angle scan of the N 1s spectrum at 565 eV photon energy,
as well as a PhD photon energy scan between 434 and 834 eV at normal emission.
At the resulting low kinetic energy of the excited electrons, back-scattering from the Rh substrate is enhanced,
and the interference pattern of the scattered photoelectron wave with the direct wave becomes sensitive to the adsorption geometry.
The resulting measurements are displayed in Fig.~\ref{figxpd} separately for the pore and wire components.
It is immediately evident, that the XPD pattern of the pore peak (panel a) exhibits more details near the normal emission direction.
Due to the shorter distance and approximate top geometry of N atoms on Rh in the pore region the interference is stronger.
In contrast, the pattern of the wire peak (panel e) is dominated by scattering within h-BN, which gives rise to the broad features at high emission angles.

For a quantitative analysis
we compare the XPD/PhD modulation patterns to multiple scattering cluster calculations based on the EDAC code \cite{Abajo2001}. 
Since photoelectron diffraction probes a very local environment,
we approximate the pore and wire regions separately with a flat h-BN layer model on top of three Rh layers.
Each cluster is 20 \AA{} wide and contains 160 atoms in total.
We expect the diffraction signal to be strongest from the three local high-symmetry positions shown in Fig.~\ref{figxpd}(j) and assign them to the pore and wire peaks as follows \cite{Hofmann1994}.
In the center of the pore (white dot in the STM image of Fig.~\ref{fig1}(a),
N occupies the {\it top} site, i.e., 
there is a Rh atom in the first layer of the substrate directly below the emitter N, 
and B occupies the {\it fcc} three-fold hollow site, i.e., (N,B) = ({\it top},{\it fcc}). 
In the wire region,
the local registry at the positions of the black dots in Fig.~\ref{fig1}(a) is different.
At one it is (N,B) = ({\it hcp},{\it top}) and at the other ({\it fcc},{\it hcp}). 
The pattern of the pore is, thus, given directly by the calculation of the ({\it top},{\it fcc}) geometry,
while the modulation of the wire peak is the sum of the ({\it hcp},{\it top}) and ({\it fcc},{\it hcp}) patterns.
Moreover, our sample contains a h-BN twin domain rotated by $180\degree$. 
The concentration of the minority domain can be determined in the same analysis of the XPD patterns and amounts to 30\%.

The best fitting structure is determined by variation of parameters to find the minimum of a reliability factor (R-factor) \cite{MuntwilerM17JSR}. 
The search algorithm is based on the efficient {\it particle swarm optimization} (PSO)
for finding the global optimum in a multi-dimensional parameter space \cite{Duncan2012}.
In our case, the parameter space is spanned by the main structural parameters (atomic distances) $d_{\mathrm{NRh}}$ and $d_{\mathrm{NB}}$ (cf. Fig.~\ref{figxpd}(i)),
as well as non-structural parameters such as the inner potential, position of the refractive surface, Debye temperature and maximum scattering path length.
$d_{\mathrm{NRh}}$ is the vertical distance of the N sublattice to the top Rh plane, 
and $d_{\mathrm{NB}}$ is the vertical distance of the N and B sublattices (buckling) of the h-BN layer.
The XPD and PhD experiments are simulated and relaxed independently. 
The best-fit simulations are shown in Fig.~\ref{figxpd}.
The R-factor curves in panels (d) and (h) are obtained after the relaxation process, 
stepping only one selected structural parameter around its optimum. 
The width of the curves at the $R = 1.16 R_{\mathrm{min}}$ level for XPD and at the $R = 1.36 R_{\mathrm{min}}$ level for PhD is used to estimate the error of the respective parameter (see supporting information).
The resulting parameters are listed in Table \ref{tab2}
and will be discussed in detail next.

\begin{table}[]
	\caption{\label{tab2}%
		Results obtained from fitting the R-factor curves. 
		The best fit $d_0$ and error estimate $\delta d$ of a parameter is determined
		from the minimum $R_{\mathrm{min}}$ and the curvature of the R-factor curve.
		The assignment of the parameters $d_{\mathrm{NRh}}$ and $d_{\mathrm{NB}}$ is shown in Fig.~\ref{figxpd}(i).
	}
	\centering
	\begin{tabular}{lllccc}
		\hline
		\hline
		 &  & & $R_{\mathrm{min}}$ & $d_0$ (\AA) & $\delta d$ (\AA) \\ \hline
		$d_{\mathrm{NRh}}$ & pore & XPD & 0.11 & 2.19 & 0.06 \\
		     &      & PhD & 0.23 & 2.20 & 0.03 \\
		     &      & combined & & 2.20 & 0.03 \\
		     & wire & XPD & 0.10 & 2.94 & 0.17 \\
		     &      & PhD & 0.42 & 3.03 & 0.05 \\ 
		     &      & combined & & 2.98 & 0.09 \\
		\hline
		$d_{\mathrm{NB}}$  & pore & XPD & 0.11 & 0.07 & 0.10 \\
		     & wire & XPD & 0.10 & 0.01 & 0.16 \\
		\hline
		\hline
	\end{tabular}
\end{table}

Calculations from the best fit model for the pore peak are shown in Fig.~\ref{figxpd}, top row.
Visually, the experimental and simulated patterns match very well for both the angle scan (panels a and b) and the energy scan (panel c).
The optimized R-factor is 0.11 for XPD and 0.23 for PhD, which are both very satisfactory.
Furthermore, as seen in panel (d), the optima of both scan modes lie close together,
which means that both are sensitive to the adsorption height,
though the uncertainty is smaller for the energy scan.
Combining the two results, the distance between the N lattice and the Rh substrate in the pore regions is $d_{\mathrm{NRh,pore}} = (2.20 \pm 0.03) \,\mbox{\AA}$.

The middle row of Fig.~\ref{figxpd} shows the analogous results for the wire peak.
The structural optimizations result in an excellent R-factor of 0.10 for the angle scan,
and in a higher R-factor of 0.42 for the energy scan.
Despite the higher R-factor, 
the major features of the PhD modulation curve in panel (g) are reproduced,
and the parabolic R-factor curve in panel (h) is much sharper than the one of the XPD scans.
The optimum values still lie close,
and the combined, weighted result for the distance between the N lattice and the Rh substrate in the wire region is $d_{\mathrm{NRh,wire}} = (2.98 \pm 0.09) \,\mbox{\AA}$.
Based on the results obtained for the pore and wire regions,
we obtain the corrugation of the h-BN/Rh(111) nanomesh $(0.8 \pm 0.1) \,\mbox{\AA}$.

Our results on the pore region compare well to experimental studies of other strongly bound h-BN systems.
On Ir(111) the adsorption distance was determined as 2.22 \AA{} by XSW \cite{Hagen2016},
on Ni(111) as 2.11 \AA{} by PhD \cite{MuntwilerM17JSR},
and on Co(0001) as 2.11 \AA{} by holographic XPD \cite{Usachov2018}.
For the weakly bound h-BN areas on Ir(111) a distance of 3.72 \AA{} was reported from XSW \cite{Hagen2016}.
On Cu(111),
where the h-BN layer is weakly bound in the whole supercell and the corrugation is much smaller,
two independent XSW studies report 3.37 \AA{} \cite{SchwarzM17AN} and 3.22 \AA{} \cite{Brulke2017}.
Since the weakly bound regions take up most of the strain from the system-dependent lattice mismatch,
it is expected that the corrugation may vary significantly between different substrates.

For h-BN/Rh(111), previous results from DFT calculations and STM/AFM measurements are available for comparison, cf. Table \ref{tab1}.
DFT results fall roughly into two classes depending on whether van der Waals (vdW) interactions are included or not.
Calculations without vdW corrections come close to the 2.20 \AA{} adsorption height measured here, 
but significantly overestimate the corrugation by a factor 2 or more.
WC-GGA and revPBE-D3 give the closest match.
The vdW-DF functional yields a better match of the corrugation, 
but overestimates the adsorption distance in the strongly bound regions.
Furthermore, vdW-DF fails to predict
the lateral size of the pores as seen in a decoration experiment \cite{Iannuzzi2014}.
From the scanning probe measurements, no decisive result can be derived due to a large and uniform spread of results.
The closest result to our findings comes from non-contact atomic force microscopy (AFM).
Though it may be less sensitive to differences in valence states,
the AFM signal is ultimately governed by the interaction with the electron density.
Thus, it is {\it a priori} not clear whether AFM can provide the same accuracy on other systems.

Discussing the {\em true} corrugation we have to bear in mind that experimental values from spatially averaging techniques represent an ensemble average of atomic positions.
In the present study the size of the averaging region is limited by two effects:
First, strongly and weakly bound regions produce two XPS peaks that can be distinguished by their binding energy.
Second, as explained in the SI, the modulation of the XPD signal is strongest around high-symmetry sites.
This is a result of the short distance between emitter an scatterer
as well as the cancelling effect of evenly distributed, low-symmetry emitter-scatterer geometries.
This renders XPD measurements sensitive to a small region of two to three h-BN lattice cells in diameter around the high-symmetry sites ({\it top},{\it fcc}) in the center of the pore, 
({\it hcp},{\it top}) and ({\it fcc},{\it hcp}) on the wire network.
From this, we infer that our results are, within the error margins, close to the actual maximum-minimum corrugation of the nanomesh.
On the other hand, PED is blind to the transition regions between wire and pore regions.
Hence, we cannot comment on the smoothness or sharpness of the transition regions.
Furthermore, the simulated pattern is more sensitive to the ({\it hcp},{\it top}) site and is not able to distinguish the two wire sites.

\begin{figure}[ht]
	\centering
	\includegraphics[scale=1]{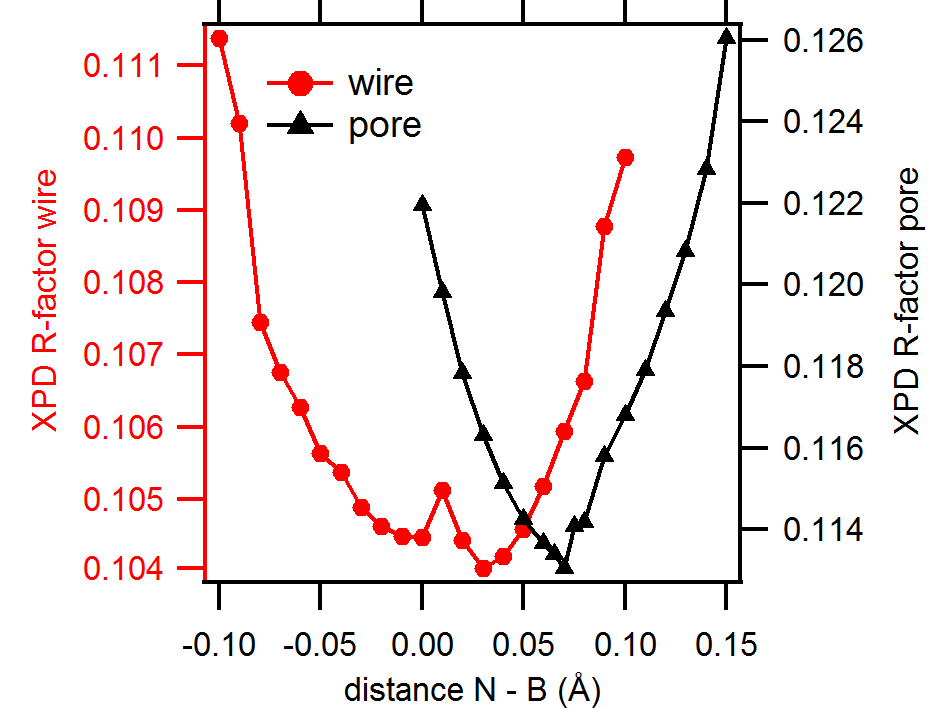}
	\caption{\label{fig4}%
		N-B buckling: XPD R-factor as a function of the distance $d_{\mathrm{NB}}$ (see Fig.~\ref{figxpd}(i))
		in the pore (red dots) and wire (black triangles) regions.}
\end{figure}

As a final point, we discuss the {\it buckling}, i.e., the vertical displacement $d_{\mathrm{NB}}$ of the N and B sublattices as illustrated in Fig.~\ref{figxpd}(i).
Photoelectron diffraction has been successfully used to quantify the buckling in other 2D systems, 
for instance, graphene on SiC \cite{deLima2013,deLima2014} and black phosphorus \cite{deLima2016}. 
The presence of buckling may have an important effect on the stability of these type of single-layer systems \cite{Sclauzero2012,deLima2013}. 
Fig.~\ref{fig4} shows the R-factor as a function of  $d_{\mathrm{NB}}$, 
where positive values mean that the B sublattice lies closer to the Rh substrate. 
The results displayed are only for XPD, 
since the PhD scan at normal emission is not sensitive to this parameter due to a very small scattering factor. 
Though it is possible to observe very clear minima in the curves of Fig.~\ref{fig4}, 
the results include a considerable uncertainty due to the low dependence of the R-factor on $d_{\mathrm{NB}}$.
The results obtained are $d_{\mathrm{NB}} = (0.07 \pm 0.10) \mbox{\AA}$ for the pores 
and $(0.01 \pm 0.16) \mbox{\AA}$ for the wires. 
Despite the large uncertainty, 
the measurement shows a clear difference between pore and wire regions.
The buckling has a significant non-zero amplitude in the strongly bound pore regions,
but is absent in the weakly bound wire regions.
In agreement with DFT results the B sub-lattice lies closer to the substrate than the N sub-lattice.
Laskowski {\it et al.} report a larger buckling of 0.14 \AA{} in the pore region \cite{Laskowski2007}. 

In conclusion, we have measured the adsorbate-substrate distance and corrugation of the h-BN nanomesh
with chemically resolved, angle- and energy-scanned photoelectron diffraction,
a method that is inherently sensitive to the position of the atomic cores. 
The combination of angle and energy scanned measurements is an advanced experimental scheme
that can reveal complementary information or provide more accurate values.
The inherent sensitivity to high-symmetry sites limits the size of the ensemble-average and yields results that are close to the true minimum-maximum atomic positions.
Our findings are important for the understanding of the adsorption behaviour of weakly bound two-dimensional layers on metal adsorbates.
To date, DFT is still not predictive in both the strongly and weakly bound regions of the layer at the same time in an {\it ab initio} way.
This stresses the fact that accurate experimental input is required to strengthen theoretical models and -- as a general objective -- to improve theoretical methods.
On the experimental side, care has to be taken as well.
While there have been several reports from common scanning probe studies,
they did not establish a reliable, reproducible corrugation value due to intrinsic limitations of probing the electronic density of states rather than the position of the atomic cores.
X-ray based methods, including XPD, require sophisticated simulations to extract the final structural parameters.
The complexity and approximations in these simulations suggest
that results should be cross-checked with different methods.
Considering prospective applications of 2D materials in novel electronic and spintronic devices
that are governed by weak van-der-Waals coupling in one region and stronger covalent bonding in another part,
it seems important that experimental tools are established
which provide non-ambiguous, quantitative details of the local atomic structure and layer distances at their interface
as an input for the development of theoretical methods that provide the necessary predictive power needed for materials engineering.

\begin{acknowledgments}

L.H.L would like to thank the CNPq-Brazil (201932/2015-6) and FAPESP-Brazil (2016/21402-8) for financial support and the hospitality of the University of Z\"{u}rich and Paul Scherrer Institut.

\end{acknowledgments}

\bibliography{references}

\begin{thebibliography}{55}
\expandafter\ifx\csname natexlab\endcsname\relax\def\natexlab#1{#1}\fi
\expandafter\ifx\csname bibnamefont\endcsname\relax
  \def\bibnamefont#1{#1}\fi
\expandafter\ifx\csname bibfnamefont\endcsname\relax
  \def\bibfnamefont#1{#1}\fi
\expandafter\ifx\csname citenamefont\endcsname\relax
  \def\citenamefont#1{#1}\fi
\expandafter\ifx\csname url\endcsname\relax
  \def\url#1{\texttt{#1}}\fi
\expandafter\ifx\csname urlprefix\endcsname\relax\def\urlprefix{URL }\fi
\providecommand{\bibinfo}[2]{#2}
\providecommand{\eprint}[2][]{\url{#2}}

\bibitem[{\citenamefont{Novoselov et~al.}(2004)\citenamefont{Novoselov, Geim,
  Morozov, Jiang, Zhang, Dubonos, Grigorieva, and Firsov}}]{Novoselov2004}
\bibinfo{author}{\bibfnamefont{K.~S.} \bibnamefont{Novoselov}},
  \bibinfo{author}{\bibfnamefont{A.~K.} \bibnamefont{Geim}},
  \bibinfo{author}{\bibfnamefont{S.~V.} \bibnamefont{Morozov}},
  \bibinfo{author}{\bibfnamefont{D.}~\bibnamefont{Jiang}},
  \bibinfo{author}{\bibfnamefont{Y.}~\bibnamefont{Zhang}},
  \bibinfo{author}{\bibfnamefont{S.~V.} \bibnamefont{Dubonos}},
  \bibinfo{author}{\bibfnamefont{I.~V.} \bibnamefont{Grigorieva}},
  \bibnamefont{and} \bibinfo{author}{\bibfnamefont{A.~A.}
  \bibnamefont{Firsov}}, \bibinfo{journal}{Science}
  \textbf{\bibinfo{volume}{306}}, \bibinfo{pages}{666} (\bibinfo{year}{2004}),
  ISSN \bibinfo{issn}{0036-8075},
  \eprint{https://science.sciencemag.org/content/306/5696/666.full.pdf},
  \urlprefix\url{https://science.sciencemag.org/content/306/5696/666}.

\bibitem[{\citenamefont{Li et~al.}(2014)\citenamefont{Li, Yu, Ye, Ge, Ou, Wu,
  Feng, Chen, and Zhang}}]{LiL14NN}
\bibinfo{author}{\bibfnamefont{L.}~\bibnamefont{Li}},
  \bibinfo{author}{\bibfnamefont{Y.}~\bibnamefont{Yu}},
  \bibinfo{author}{\bibfnamefont{G.~J.} \bibnamefont{Ye}},
  \bibinfo{author}{\bibfnamefont{Q.}~\bibnamefont{Ge}},
  \bibinfo{author}{\bibfnamefont{X.}~\bibnamefont{Ou}},
  \bibinfo{author}{\bibfnamefont{H.}~\bibnamefont{Wu}},
  \bibinfo{author}{\bibfnamefont{D.}~\bibnamefont{Feng}},
  \bibinfo{author}{\bibfnamefont{X.~H.} \bibnamefont{Chen}}, \bibnamefont{and}
  \bibinfo{author}{\bibfnamefont{Y.}~\bibnamefont{Zhang}},
  \bibinfo{journal}{Nat. Nanotechnol.} \textbf{\bibinfo{volume}{9}},
  \bibinfo{pages}{372} (\bibinfo{year}{2014}).

\bibitem[{\citenamefont{Nagashima et~al.}(1995)\citenamefont{Nagashima, Tejima,
  Gamou, Kawai, and Oshima}}]{Nagashima1995}
\bibinfo{author}{\bibfnamefont{A.}~\bibnamefont{Nagashima}},
  \bibinfo{author}{\bibfnamefont{N.}~\bibnamefont{Tejima}},
  \bibinfo{author}{\bibfnamefont{Y.}~\bibnamefont{Gamou}},
  \bibinfo{author}{\bibfnamefont{T.}~\bibnamefont{Kawai}}, \bibnamefont{and}
  \bibinfo{author}{\bibfnamefont{C.}~\bibnamefont{Oshima}},
  \bibinfo{journal}{Phys. Rev. B} \textbf{\bibinfo{volume}{51}},
  \bibinfo{pages}{4606} (\bibinfo{year}{1995}),
  \urlprefix\url{https://link.aps.org/doi/10.1103/PhysRevB.51.4606}.

\bibitem[{\citenamefont{Manzeli et~al.}(2017)\citenamefont{Manzeli,
  Ovchinnikov, Pasquier, Yazyev, and Kis}}]{Manzeli2017}
\bibinfo{author}{\bibfnamefont{S.}~\bibnamefont{Manzeli}},
  \bibinfo{author}{\bibfnamefont{D.}~\bibnamefont{Ovchinnikov}},
  \bibinfo{author}{\bibfnamefont{D.}~\bibnamefont{Pasquier}},
  \bibinfo{author}{\bibfnamefont{O.~V.} \bibnamefont{Yazyev}},
  \bibnamefont{and} \bibinfo{author}{\bibfnamefont{A.}~\bibnamefont{Kis}},
  \bibinfo{journal}{Nature Reviews Materials} \textbf{\bibinfo{volume}{2}},
  \bibinfo{pages}{17033} (\bibinfo{year}{2017}),
  \urlprefix\url{https://doi.org/10.1038/natrevmats.2017.33}.

\bibitem[{\citenamefont{Cao et~al.}(2018)\citenamefont{Cao, Fatemi, Fang,
  Watanabe, Taniguchi, Kaxiras, and Jarillo-Herrero}}]{CaoY18N}
\bibinfo{author}{\bibfnamefont{Y.}~\bibnamefont{Cao}},
  \bibinfo{author}{\bibfnamefont{V.}~\bibnamefont{Fatemi}},
  \bibinfo{author}{\bibfnamefont{S.}~\bibnamefont{Fang}},
  \bibinfo{author}{\bibfnamefont{K.}~\bibnamefont{Watanabe}},
  \bibinfo{author}{\bibfnamefont{T.}~\bibnamefont{Taniguchi}},
  \bibinfo{author}{\bibfnamefont{E.}~\bibnamefont{Kaxiras}}, \bibnamefont{and}
  \bibinfo{author}{\bibfnamefont{P.}~\bibnamefont{Jarillo-Herrero}},
  \bibinfo{journal}{Nature} \textbf{\bibinfo{volume}{556}}, \bibinfo{pages}{43}
  (\bibinfo{year}{2018}).

\bibitem[{\citenamefont{Wang et~al.}(2011)\citenamefont{Wang, Taychatanapat,
  Hsu, Watanabe, Taniguchi, Jarillo-Herrero, and Palacios}}]{Wang2011}
\bibinfo{author}{\bibfnamefont{H.}~\bibnamefont{Wang}},
  \bibinfo{author}{\bibfnamefont{T.}~\bibnamefont{Taychatanapat}},
  \bibinfo{author}{\bibfnamefont{A.}~\bibnamefont{Hsu}},
  \bibinfo{author}{\bibfnamefont{K.}~\bibnamefont{Watanabe}},
  \bibinfo{author}{\bibfnamefont{T.}~\bibnamefont{Taniguchi}},
  \bibinfo{author}{\bibfnamefont{P.}~\bibnamefont{Jarillo-Herrero}},
  \bibnamefont{and} \bibinfo{author}{\bibfnamefont{T.}~\bibnamefont{Palacios}},
  \bibinfo{journal}{IEEE Electron Device Letters}
  \textbf{\bibinfo{volume}{32}}, \bibinfo{pages}{1209} (\bibinfo{year}{2011}).

\bibitem[{\citenamefont{Xu et~al.}(2013)\citenamefont{Xu, Liang, Shi, and
  Chen}}]{Xu2013}
\bibinfo{author}{\bibfnamefont{M.}~\bibnamefont{Xu}},
  \bibinfo{author}{\bibfnamefont{T.}~\bibnamefont{Liang}},
  \bibinfo{author}{\bibfnamefont{M.}~\bibnamefont{Shi}}, \bibnamefont{and}
  \bibinfo{author}{\bibfnamefont{H.}~\bibnamefont{Chen}},
  \bibinfo{journal}{Chem. Rev.} \textbf{\bibinfo{volume}{113}},
  \bibinfo{pages}{3766} (\bibinfo{year}{2013}),
  \urlprefix\url{http://dx.doi.org/10.1021/cr300263a}.

\bibitem[{\citenamefont{Iannaccone et~al.}(2018)\citenamefont{Iannaccone,
  Bonaccorso, Colombo, and Fiori}}]{IannacconeG18NN}
\bibinfo{author}{\bibfnamefont{G.}~\bibnamefont{Iannaccone}},
  \bibinfo{author}{\bibfnamefont{F.}~\bibnamefont{Bonaccorso}},
  \bibinfo{author}{\bibfnamefont{L.}~\bibnamefont{Colombo}}, \bibnamefont{and}
  \bibinfo{author}{\bibfnamefont{G.}~\bibnamefont{Fiori}},
  \bibinfo{journal}{Nat. Nanotechnol.} \textbf{\bibinfo{volume}{13}},
  \bibinfo{pages}{183} (\bibinfo{year}{2018}), ISSN \bibinfo{issn}{1748-3395}.

\bibitem[{\citenamefont{Liu et~al.}(2019)\citenamefont{Liu, Huang, and
  Duan}}]{LiuY19N}
\bibinfo{author}{\bibfnamefont{Y.}~\bibnamefont{Liu}},
  \bibinfo{author}{\bibfnamefont{Y.}~\bibnamefont{Huang}}, \bibnamefont{and}
  \bibinfo{author}{\bibfnamefont{X.}~\bibnamefont{Duan}},
  \bibinfo{journal}{Nature} \textbf{\bibinfo{volume}{567}},
  \bibinfo{pages}{323} (\bibinfo{year}{2019}), ISSN \bibinfo{issn}{1476-4687}.

\bibitem[{\citenamefont{Dean et~al.}(2010)\citenamefont{Dean, Young, Meric,
  Lee, Wang, Sorgenfrei, Watanabe, Taniguchi, Kim, Shepard et~al.}}]{Dean2010}
\bibinfo{author}{\bibfnamefont{C.~R.} \bibnamefont{Dean}},
  \bibinfo{author}{\bibfnamefont{A.~F.} \bibnamefont{Young}},
  \bibinfo{author}{\bibfnamefont{I.}~\bibnamefont{Meric}},
  \bibinfo{author}{\bibfnamefont{C.}~\bibnamefont{Lee}},
  \bibinfo{author}{\bibfnamefont{L.}~\bibnamefont{Wang}},
  \bibinfo{author}{\bibfnamefont{S.}~\bibnamefont{Sorgenfrei}},
  \bibinfo{author}{\bibfnamefont{K.}~\bibnamefont{Watanabe}},
  \bibinfo{author}{\bibfnamefont{T.}~\bibnamefont{Taniguchi}},
  \bibinfo{author}{\bibfnamefont{P.}~\bibnamefont{Kim}},
  \bibinfo{author}{\bibfnamefont{K.~L.} \bibnamefont{Shepard}},
  \bibnamefont{et~al.}, \bibinfo{journal}{Nat. Nanotechnol.}
  \textbf{\bibinfo{volume}{5}}, \bibinfo{pages}{722} (\bibinfo{year}{2010}).

\bibitem[{\citenamefont{Gehring et~al.}(2012)\citenamefont{Gehring, Gao,
  Burghard, and Kern}}]{Gehring2012}
\bibinfo{author}{\bibfnamefont{P.}~\bibnamefont{Gehring}},
  \bibinfo{author}{\bibfnamefont{B.~F.} \bibnamefont{Gao}},
  \bibinfo{author}{\bibfnamefont{M.}~\bibnamefont{Burghard}}, \bibnamefont{and}
  \bibinfo{author}{\bibfnamefont{K.}~\bibnamefont{Kern}},
  \bibinfo{journal}{Nano Lett.} \textbf{\bibinfo{volume}{12}},
  \bibinfo{pages}{5137} (\bibinfo{year}{2012}).

\bibitem[{\citenamefont{Zhang et~al.}(2015)\citenamefont{Zhang, Zhu, Wang,
  Feng, Qiao, Wen, Chen, Cui, Zhang, Cai et~al.}}]{Zhang2015}
\bibinfo{author}{\bibfnamefont{M.}~\bibnamefont{Zhang}},
  \bibinfo{author}{\bibfnamefont{Y.}~\bibnamefont{Zhu}},
  \bibinfo{author}{\bibfnamefont{X.}~\bibnamefont{Wang}},
  \bibinfo{author}{\bibfnamefont{Q.}~\bibnamefont{Feng}},
  \bibinfo{author}{\bibfnamefont{S.}~\bibnamefont{Qiao}},
  \bibinfo{author}{\bibfnamefont{W.}~\bibnamefont{Wen}},
  \bibinfo{author}{\bibfnamefont{Y.}~\bibnamefont{Chen}},
  \bibinfo{author}{\bibfnamefont{M.}~\bibnamefont{Cui}},
  \bibinfo{author}{\bibfnamefont{J.}~\bibnamefont{Zhang}},
  \bibinfo{author}{\bibfnamefont{C.}~\bibnamefont{Cai}}, \bibnamefont{et~al.},
  \bibinfo{journal}{J. Am. Chem. Soc.} \textbf{\bibinfo{volume}{137}},
  \bibinfo{pages}{7051} (\bibinfo{year}{2015}).

\bibitem[{\citenamefont{Yan et~al.}(2015)\citenamefont{Yan, Velasco, Kahn,
  Watanabe, Taniguchi, Wang, Crommie, and Zettl}}]{Yan2015}
\bibinfo{author}{\bibfnamefont{A.}~\bibnamefont{Yan}},
  \bibinfo{author}{\bibfnamefont{J.}~\bibnamefont{Velasco}},
  \bibinfo{author}{\bibfnamefont{S.}~\bibnamefont{Kahn}},
  \bibinfo{author}{\bibfnamefont{K.}~\bibnamefont{Watanabe}},
  \bibinfo{author}{\bibfnamefont{T.}~\bibnamefont{Taniguchi}},
  \bibinfo{author}{\bibfnamefont{F.}~\bibnamefont{Wang}},
  \bibinfo{author}{\bibfnamefont{M.~F.} \bibnamefont{Crommie}},
  \bibnamefont{and} \bibinfo{author}{\bibfnamefont{A.}~\bibnamefont{Zettl}},
  \bibinfo{journal}{Nano Lett.} \textbf{\bibinfo{volume}{15}},
  \bibinfo{pages}{6324} (\bibinfo{year}{2015}).

\bibitem[{\citenamefont{Zhang et~al.}(2016)\citenamefont{Zhang, Chen, Zhang,
  Pan, Chou, Zeng, and Shih}}]{Zhang2016}
\bibinfo{author}{\bibfnamefont{Q.}~\bibnamefont{Zhang}},
  \bibinfo{author}{\bibfnamefont{Y.}~\bibnamefont{Chen}},
  \bibinfo{author}{\bibfnamefont{C.}~\bibnamefont{Zhang}},
  \bibinfo{author}{\bibfnamefont{C.-R.} \bibnamefont{Pan}},
  \bibinfo{author}{\bibfnamefont{M.-Y.} \bibnamefont{Chou}},
  \bibinfo{author}{\bibfnamefont{C.}~\bibnamefont{Zeng}}, \bibnamefont{and}
  \bibinfo{author}{\bibfnamefont{C.-K.} \bibnamefont{Shih}},
  \bibinfo{journal}{Nat. Commun.} \textbf{\bibinfo{volume}{7}},
  \bibinfo{pages}{13843} (\bibinfo{year}{2016}).

\bibitem[{\citenamefont{Chen et~al.}(2018)\citenamefont{Chen, Kim, Bernard,
  Pizzochero, Zaldívar, Pascual, Ugeda, Yazyev, Greber, Osterwalder
  et~al.}}]{Chen2018}
\bibinfo{author}{\bibfnamefont{M.-W.} \bibnamefont{Chen}},
  \bibinfo{author}{\bibfnamefont{H.}~\bibnamefont{Kim}},
  \bibinfo{author}{\bibfnamefont{C.}~\bibnamefont{Bernard}},
  \bibinfo{author}{\bibfnamefont{M.}~\bibnamefont{Pizzochero}},
  \bibinfo{author}{\bibfnamefont{J.}~\bibnamefont{Zaldívar}},
  \bibinfo{author}{\bibfnamefont{J.~I.} \bibnamefont{Pascual}},
  \bibinfo{author}{\bibfnamefont{M.~M.} \bibnamefont{Ugeda}},
  \bibinfo{author}{\bibfnamefont{O.~V.} \bibnamefont{Yazyev}},
  \bibinfo{author}{\bibfnamefont{T.}~\bibnamefont{Greber}},
  \bibinfo{author}{\bibfnamefont{J.}~\bibnamefont{Osterwalder}},
  \bibnamefont{et~al.}, \bibinfo{journal}{ACS Nano}
  \textbf{\bibinfo{volume}{12}}, \bibinfo{pages}{11161} (\bibinfo{year}{2018}),
  \bibinfo{note}{pMID: 30371049},
  \eprint{https://doi.org/10.1021/acsnano.8b05628},
  \urlprefix\url{https://doi.org/10.1021/acsnano.8b05628}.

\bibitem[{\citenamefont{Auwärter}(2019)}]{AuwarterW18SSR}
\bibinfo{author}{\bibfnamefont{W.}~\bibnamefont{Auwärter}},
  \bibinfo{journal}{Surf. Sci. Rep.} \textbf{\bibinfo{volume}{74}},
  \bibinfo{pages}{1 } (\bibinfo{year}{2019}), ISSN \bibinfo{issn}{0167-5729},
  \urlprefix\url{http://www.sciencedirect.com/science/article/pii/S0167572918300517}.

\bibitem[{\citenamefont{Preobrajenski et~al.}(2007)\citenamefont{Preobrajenski,
  Nesterov, Ng, Vinogradov, and M\aa{}rtensson}}]{Preobrajenski2007}
\bibinfo{author}{\bibfnamefont{A.}~\bibnamefont{Preobrajenski}},
  \bibinfo{author}{\bibfnamefont{M.}~\bibnamefont{Nesterov}},
  \bibinfo{author}{\bibfnamefont{M.~L.} \bibnamefont{Ng}},
  \bibinfo{author}{\bibfnamefont{A.}~\bibnamefont{Vinogradov}},
  \bibnamefont{and}
  \bibinfo{author}{\bibfnamefont{N.}~\bibnamefont{M\aa{}rtensson}},
  \bibinfo{journal}{Chem. Phys. Lett.} \textbf{\bibinfo{volume}{446}},
  \bibinfo{pages}{119 } (\bibinfo{year}{2007}), ISSN \bibinfo{issn}{0009-2614},
  \urlprefix\url{http://www.sciencedirect.com/science/article/pii/S0009261407010998}.

\bibitem[{\citenamefont{Brugger et~al.}(2009)\citenamefont{Brugger, G\"unther,
  Wang, Dil, Bocquet, Osterwalder, Wintterlin, and Greber}}]{Brugger2007}
\bibinfo{author}{\bibfnamefont{T.}~\bibnamefont{Brugger}},
  \bibinfo{author}{\bibfnamefont{S.}~\bibnamefont{G\"unther}},
  \bibinfo{author}{\bibfnamefont{B.}~\bibnamefont{Wang}},
  \bibinfo{author}{\bibfnamefont{J.~H.} \bibnamefont{Dil}},
  \bibinfo{author}{\bibfnamefont{M.-L.} \bibnamefont{Bocquet}},
  \bibinfo{author}{\bibfnamefont{J.}~\bibnamefont{Osterwalder}},
  \bibinfo{author}{\bibfnamefont{J.}~\bibnamefont{Wintterlin}},
  \bibnamefont{and} \bibinfo{author}{\bibfnamefont{T.}~\bibnamefont{Greber}},
  \bibinfo{journal}{Phys. Rev. B} \textbf{\bibinfo{volume}{79}},
  \bibinfo{pages}{045407} (\bibinfo{year}{2009}),
  \urlprefix\url{http://link.aps.org/doi/10.1103/PhysRevB.79.045407}.

\bibitem[{\citenamefont{Preobrajenski et~al.}(2008)\citenamefont{Preobrajenski,
  Ng, Vinogradov, and M\aa{}rtensson}}]{Preobrajenski2008}
\bibinfo{author}{\bibfnamefont{A.~B.} \bibnamefont{Preobrajenski}},
  \bibinfo{author}{\bibfnamefont{M.~L.} \bibnamefont{Ng}},
  \bibinfo{author}{\bibfnamefont{A.~S.} \bibnamefont{Vinogradov}},
  \bibnamefont{and}
  \bibinfo{author}{\bibfnamefont{N.}~\bibnamefont{M\aa{}rtensson}},
  \bibinfo{journal}{Phys. Rev. B} \textbf{\bibinfo{volume}{78}},
  \bibinfo{pages}{073401} (\bibinfo{year}{2008}),
  \urlprefix\url{http://link.aps.org/doi/10.1103/PhysRevB.78.073401}.

\bibitem[{\citenamefont{Corso et~al.}(2004)\citenamefont{Corso, Auw{\"a}rter,
  Muntwiler, Tamai, Greber, and Osterwalder}}]{Corso2004}
\bibinfo{author}{\bibfnamefont{M.}~\bibnamefont{Corso}},
  \bibinfo{author}{\bibfnamefont{W.}~\bibnamefont{Auw{\"a}rter}},
  \bibinfo{author}{\bibfnamefont{M.}~\bibnamefont{Muntwiler}},
  \bibinfo{author}{\bibfnamefont{A.}~\bibnamefont{Tamai}},
  \bibinfo{author}{\bibfnamefont{T.}~\bibnamefont{Greber}}, \bibnamefont{and}
  \bibinfo{author}{\bibfnamefont{J.}~\bibnamefont{Osterwalder}},
  \bibinfo{journal}{Science} \textbf{\bibinfo{volume}{303}},
  \bibinfo{pages}{217} (\bibinfo{year}{2004}), ISSN \bibinfo{issn}{0036-8075},
  \urlprefix\url{http://science.sciencemag.org/content/303/5655/217}.

\bibitem[{\citenamefont{Müller et~al.}(2009)\citenamefont{Müller, Sachdev,
  Hüfner, Pollard, Perkins, Russell, Beton, Gsell, Fischer, Schreck
  et~al.}}]{Muller2009}
\bibinfo{author}{\bibfnamefont{F.}~\bibnamefont{Müller}},
  \bibinfo{author}{\bibfnamefont{H.}~\bibnamefont{Sachdev}},
  \bibinfo{author}{\bibfnamefont{S.}~\bibnamefont{Hüfner}},
  \bibinfo{author}{\bibfnamefont{A.~J.} \bibnamefont{Pollard}},
  \bibinfo{author}{\bibfnamefont{E.~W.} \bibnamefont{Perkins}},
  \bibinfo{author}{\bibfnamefont{J.~C.} \bibnamefont{Russell}},
  \bibinfo{author}{\bibfnamefont{P.~H.} \bibnamefont{Beton}},
  \bibinfo{author}{\bibfnamefont{S.}~\bibnamefont{Gsell}},
  \bibinfo{author}{\bibfnamefont{M.}~\bibnamefont{Fischer}},
  \bibinfo{author}{\bibfnamefont{M.}~\bibnamefont{Schreck}},
  \bibnamefont{et~al.}, \bibinfo{journal}{Small} \textbf{\bibinfo{volume}{5}},
  \bibinfo{pages}{2291} (\bibinfo{year}{2009}),
  \urlprefix\url{http://dx.doi.org/10.1002/smll.200900158}.

\bibitem[{\citenamefont{Martoccia
  et~al.}(2010{\natexlab{a}})\citenamefont{Martoccia, Pauli, Brugger, Greber,
  Patterson, and Willmott}}]{Martoccia2010_Rh111}
\bibinfo{author}{\bibfnamefont{D.}~\bibnamefont{Martoccia}},
  \bibinfo{author}{\bibfnamefont{S.}~\bibnamefont{Pauli}},
  \bibinfo{author}{\bibfnamefont{T.}~\bibnamefont{Brugger}},
  \bibinfo{author}{\bibfnamefont{T.}~\bibnamefont{Greber}},
  \bibinfo{author}{\bibfnamefont{B.}~\bibnamefont{Patterson}},
  \bibnamefont{and} \bibinfo{author}{\bibfnamefont{P.}~\bibnamefont{Willmott}},
  \bibinfo{journal}{Surface Science} \textbf{\bibinfo{volume}{604}},
  \bibinfo{pages}{L9 } (\bibinfo{year}{2010}{\natexlab{a}}), ISSN
  \bibinfo{issn}{0039-6028},
  \urlprefix\url{http://www.sciencedirect.com/science/article/pii/S0039602809007936}.

\bibitem[{\citenamefont{Roth et~al.}(2011)\citenamefont{Roth, Osterwalder, and
  Greber}}]{Roth2011}
\bibinfo{author}{\bibfnamefont{S.}~\bibnamefont{Roth}},
  \bibinfo{author}{\bibfnamefont{J.}~\bibnamefont{Osterwalder}},
  \bibnamefont{and} \bibinfo{author}{\bibfnamefont{T.}~\bibnamefont{Greber}},
  \bibinfo{journal}{Surf. Sci.} \textbf{\bibinfo{volume}{605}},
  \bibinfo{pages}{L17 } (\bibinfo{year}{2011}), ISSN \bibinfo{issn}{0039-6028},
  \urlprefix\url{http://www.sciencedirect.com/science/article/pii/S0039602811000586}.

\bibitem[{\citenamefont{Paffett et~al.}(1990)\citenamefont{Paffett, Simonson,
  Papin, and Paine}}]{Paffett1990}
\bibinfo{author}{\bibfnamefont{M.}~\bibnamefont{Paffett}},
  \bibinfo{author}{\bibfnamefont{R.}~\bibnamefont{Simonson}},
  \bibinfo{author}{\bibfnamefont{P.}~\bibnamefont{Papin}}, \bibnamefont{and}
  \bibinfo{author}{\bibfnamefont{R.}~\bibnamefont{Paine}},
  \bibinfo{journal}{Surface Science} \textbf{\bibinfo{volume}{232}},
  \bibinfo{pages}{286 } (\bibinfo{year}{1990}), ISSN \bibinfo{issn}{0039-6028},
  \urlprefix\url{http://www.sciencedirect.com/science/article/pii/003960289090121N}.

\bibitem[{\citenamefont{Goriachko et~al.}(2007)\citenamefont{Goriachko, He,
  Knapp, Over, Corso, Brugger, Berner, Osterwalder, and
  Greber}}]{Goriachko2007}
\bibinfo{author}{\bibfnamefont{A.}~\bibnamefont{Goriachko}},
  \bibinfo{author}{\bibnamefont{He}},
  \bibinfo{author}{\bibfnamefont{M.}~\bibnamefont{Knapp}},
  \bibinfo{author}{\bibfnamefont{H.}~\bibnamefont{Over}},
  \bibinfo{author}{\bibfnamefont{M.}~\bibnamefont{Corso}},
  \bibinfo{author}{\bibfnamefont{T.}~\bibnamefont{Brugger}},
  \bibinfo{author}{\bibfnamefont{S.}~\bibnamefont{Berner}},
  \bibinfo{author}{\bibfnamefont{J.}~\bibnamefont{Osterwalder}},
  \bibnamefont{and} \bibinfo{author}{\bibfnamefont{T.}~\bibnamefont{Greber}},
  \bibinfo{journal}{Langmuir} \textbf{\bibinfo{volume}{23}},
  \bibinfo{pages}{2928} (\bibinfo{year}{2007}).

\bibitem[{\citenamefont{Martoccia
  et~al.}(2010{\natexlab{b}})\citenamefont{Martoccia, Brugger, Björck,
  Schlepütz, Pauli, Greber, Patterson, and Willmott}}]{Martoccia2010_Ru0001}
\bibinfo{author}{\bibfnamefont{D.}~\bibnamefont{Martoccia}},
  \bibinfo{author}{\bibfnamefont{T.}~\bibnamefont{Brugger}},
  \bibinfo{author}{\bibfnamefont{M.}~\bibnamefont{Björck}},
  \bibinfo{author}{\bibfnamefont{C.}~\bibnamefont{Schlepütz}},
  \bibinfo{author}{\bibfnamefont{S.}~\bibnamefont{Pauli}},
  \bibinfo{author}{\bibfnamefont{T.}~\bibnamefont{Greber}},
  \bibinfo{author}{\bibfnamefont{B.}~\bibnamefont{Patterson}},
  \bibnamefont{and} \bibinfo{author}{\bibfnamefont{P.}~\bibnamefont{Willmott}},
  \bibinfo{journal}{Surface Science} \textbf{\bibinfo{volume}{604}},
  \bibinfo{pages}{L16 } (\bibinfo{year}{2010}{\natexlab{b}}), ISSN
  \bibinfo{issn}{0039-6028},
  \urlprefix\url{http://www.sciencedirect.com/science/article/pii/S0039602810000051}.

\bibitem[{\citenamefont{Berner et~al.}(2007)\citenamefont{Berner, Corso,
  Widmer, Groening, Laskowski, Blaha, Schwarz, Goriachko, Over, Gsell
  et~al.}}]{Berner2007}
\bibinfo{author}{\bibfnamefont{S.}~\bibnamefont{Berner}},
  \bibinfo{author}{\bibfnamefont{M.}~\bibnamefont{Corso}},
  \bibinfo{author}{\bibfnamefont{R.}~\bibnamefont{Widmer}},
  \bibinfo{author}{\bibfnamefont{O.}~\bibnamefont{Groening}},
  \bibinfo{author}{\bibfnamefont{R.}~\bibnamefont{Laskowski}},
  \bibinfo{author}{\bibfnamefont{P.}~\bibnamefont{Blaha}},
  \bibinfo{author}{\bibfnamefont{K.}~\bibnamefont{Schwarz}},
  \bibinfo{author}{\bibfnamefont{A.}~\bibnamefont{Goriachko}},
  \bibinfo{author}{\bibfnamefont{H.}~\bibnamefont{Over}},
  \bibinfo{author}{\bibfnamefont{S.}~\bibnamefont{Gsell}},
  \bibnamefont{et~al.}, \bibinfo{journal}{Angew. Chem. Int. Ed.}
  \textbf{\bibinfo{volume}{46}}, \bibinfo{pages}{5115} (\bibinfo{year}{2007}),
  \urlprefix\url{http://dx.doi.org/10.1002/anie.200700234}.

\bibitem[{\citenamefont{Dil et~al.}(2008)\citenamefont{Dil, Lobo-Checa,
  Laskowski, Blaha, Berner, Osterwalder, and Greber}}]{Dil2008}
\bibinfo{author}{\bibfnamefont{H.}~\bibnamefont{Dil}},
  \bibinfo{author}{\bibfnamefont{J.}~\bibnamefont{Lobo-Checa}},
  \bibinfo{author}{\bibfnamefont{R.}~\bibnamefont{Laskowski}},
  \bibinfo{author}{\bibfnamefont{P.}~\bibnamefont{Blaha}},
  \bibinfo{author}{\bibfnamefont{S.}~\bibnamefont{Berner}},
  \bibinfo{author}{\bibfnamefont{J.}~\bibnamefont{Osterwalder}},
  \bibnamefont{and} \bibinfo{author}{\bibfnamefont{T.}~\bibnamefont{Greber}},
  \bibinfo{journal}{Science} \textbf{\bibinfo{volume}{319}},
  \bibinfo{pages}{1824} (\bibinfo{year}{2008}), ISSN \bibinfo{issn}{0036-8075},
  \urlprefix\url{http://science.sciencemag.org/content/319/5871/1824}.

\bibitem[{\citenamefont{Brihuega et~al.}(2008)\citenamefont{Brihuega,
  Michaelis, Zhang, Bose, Sessi, Honolka, Schneider, Enders, and
  Kern}}]{Brihuega2008}
\bibinfo{author}{\bibfnamefont{I.}~\bibnamefont{Brihuega}},
  \bibinfo{author}{\bibfnamefont{C.~H.} \bibnamefont{Michaelis}},
  \bibinfo{author}{\bibfnamefont{J.}~\bibnamefont{Zhang}},
  \bibinfo{author}{\bibfnamefont{S.}~\bibnamefont{Bose}},
  \bibinfo{author}{\bibfnamefont{V.}~\bibnamefont{Sessi}},
  \bibinfo{author}{\bibfnamefont{J.}~\bibnamefont{Honolka}},
  \bibinfo{author}{\bibfnamefont{M.~A.} \bibnamefont{Schneider}},
  \bibinfo{author}{\bibfnamefont{A.}~\bibnamefont{Enders}}, \bibnamefont{and}
  \bibinfo{author}{\bibfnamefont{K.}~\bibnamefont{Kern}},
  \bibinfo{journal}{Surf. Sci.} \textbf{\bibinfo{volume}{602}},
  \bibinfo{pages}{L95 } (\bibinfo{year}{2008}), ISSN \bibinfo{issn}{0039-6028},
  \urlprefix\url{http://www.sciencedirect.com/science/article/pii/S0039602808002951}.

\bibitem[{\citenamefont{Ma et~al.}(2010)\citenamefont{Ma, Brugger, Berner,
  Ding, Iannuzzi, Hutter, Osterwalder, and Greber}}]{Ma2010}
\bibinfo{author}{\bibfnamefont{H.}~\bibnamefont{Ma}},
  \bibinfo{author}{\bibfnamefont{T.}~\bibnamefont{Brugger}},
  \bibinfo{author}{\bibfnamefont{S.}~\bibnamefont{Berner}},
  \bibinfo{author}{\bibfnamefont{Y.}~\bibnamefont{Ding}},
  \bibinfo{author}{\bibfnamefont{M.}~\bibnamefont{Iannuzzi}},
  \bibinfo{author}{\bibfnamefont{J.}~\bibnamefont{Hutter}},
  \bibinfo{author}{\bibfnamefont{J.}~\bibnamefont{Osterwalder}},
  \bibnamefont{and} \bibinfo{author}{\bibfnamefont{T.}~\bibnamefont{Greber}},
  \bibinfo{journal}{ChemPhysChem} \textbf{\bibinfo{volume}{11}},
  \bibinfo{pages}{399} (\bibinfo{year}{2010}), ISSN \bibinfo{issn}{1439-7641},
  \urlprefix\url{http://dx.doi.org/10.1002/cphc.200900857}.

\bibitem[{\citenamefont{Ding et~al.}(2011)\citenamefont{Ding, Iannuzzi, and
  Hutter}}]{Ding2011}
\bibinfo{author}{\bibfnamefont{Y.}~\bibnamefont{Ding}},
  \bibinfo{author}{\bibfnamefont{M.}~\bibnamefont{Iannuzzi}}, \bibnamefont{and}
  \bibinfo{author}{\bibfnamefont{J.}~\bibnamefont{Hutter}},
  \bibinfo{journal}{J. Phys. Chem. C} \textbf{\bibinfo{volume}{115}},
  \bibinfo{pages}{13685} (\bibinfo{year}{2011}).

\bibitem[{\citenamefont{Patterson et~al.}(2014)\citenamefont{Patterson,
  Habenicht, Kurtz, Liu, Xu, and Sprunger}}]{Patterson2014}
\bibinfo{author}{\bibfnamefont{M.~C.} \bibnamefont{Patterson}},
  \bibinfo{author}{\bibfnamefont{B.~F.} \bibnamefont{Habenicht}},
  \bibinfo{author}{\bibfnamefont{R.~L.} \bibnamefont{Kurtz}},
  \bibinfo{author}{\bibfnamefont{L.}~\bibnamefont{Liu}},
  \bibinfo{author}{\bibfnamefont{Y.}~\bibnamefont{Xu}}, \bibnamefont{and}
  \bibinfo{author}{\bibfnamefont{P.~T.} \bibnamefont{Sprunger}},
  \bibinfo{journal}{Phys. Rev. B} \textbf{\bibinfo{volume}{89}},
  \bibinfo{pages}{205423} (\bibinfo{year}{2014}),
  \urlprefix\url{http://link.aps.org/doi/10.1103/PhysRevB.89.205423}.

\bibitem[{\citenamefont{Iannuzzi et~al.}(2014)\citenamefont{Iannuzzi, Tran,
  Widmer, Dienel, Radican, Ding, Hutter, and Groning}}]{Iannuzzi2014}
\bibinfo{author}{\bibfnamefont{M.}~\bibnamefont{Iannuzzi}},
  \bibinfo{author}{\bibfnamefont{F.}~\bibnamefont{Tran}},
  \bibinfo{author}{\bibfnamefont{R.}~\bibnamefont{Widmer}},
  \bibinfo{author}{\bibfnamefont{T.}~\bibnamefont{Dienel}},
  \bibinfo{author}{\bibfnamefont{K.}~\bibnamefont{Radican}},
  \bibinfo{author}{\bibfnamefont{Y.}~\bibnamefont{Ding}},
  \bibinfo{author}{\bibfnamefont{J.}~\bibnamefont{Hutter}}, \bibnamefont{and}
  \bibinfo{author}{\bibfnamefont{O.}~\bibnamefont{Groning}},
  \bibinfo{journal}{Phys. Chem. Chem. Phys.} \textbf{\bibinfo{volume}{16}},
  \bibinfo{pages}{12374} (\bibinfo{year}{2014}),
  \urlprefix\url{http://dx.doi.org/10.1039/C4CP01466A}.

\bibitem[{\citenamefont{Joshi et~al.}(2012)\citenamefont{Joshi, Ecija, Koitz,
  Iannuzzi, Seitsonen, Hutter, Sachdev, Vijayaraghavan, Bischoff, Seufert
  et~al.}}]{Joshi12NL}
\bibinfo{author}{\bibfnamefont{S.}~\bibnamefont{Joshi}},
  \bibinfo{author}{\bibfnamefont{D.}~\bibnamefont{Ecija}},
  \bibinfo{author}{\bibfnamefont{R.}~\bibnamefont{Koitz}},
  \bibinfo{author}{\bibfnamefont{M.}~\bibnamefont{Iannuzzi}},
  \bibinfo{author}{\bibfnamefont{A.~P.} \bibnamefont{Seitsonen}},
  \bibinfo{author}{\bibfnamefont{J.}~\bibnamefont{Hutter}},
  \bibinfo{author}{\bibfnamefont{H.}~\bibnamefont{Sachdev}},
  \bibinfo{author}{\bibfnamefont{S.}~\bibnamefont{Vijayaraghavan}},
  \bibinfo{author}{\bibfnamefont{F.}~\bibnamefont{Bischoff}},
  \bibinfo{author}{\bibfnamefont{K.}~\bibnamefont{Seufert}},
  \bibnamefont{et~al.}, \bibinfo{journal}{Nano Lett.}
  \textbf{\bibinfo{volume}{12}}, \bibinfo{pages}{5821} (\bibinfo{year}{2012}).

\bibitem[{\citenamefont{Schwarz et~al.}(2017)\citenamefont{Schwarz, Riss,
  Garnica, Ducke, Deimel, Duncan, Thakur, Lee, Seitsonen, Barth
  et~al.}}]{SchwarzM17AN}
\bibinfo{author}{\bibfnamefont{M.}~\bibnamefont{Schwarz}},
  \bibinfo{author}{\bibfnamefont{A.}~\bibnamefont{Riss}},
  \bibinfo{author}{\bibfnamefont{M.}~\bibnamefont{Garnica}},
  \bibinfo{author}{\bibfnamefont{J.}~\bibnamefont{Ducke}},
  \bibinfo{author}{\bibfnamefont{P.~S.} \bibnamefont{Deimel}},
  \bibinfo{author}{\bibfnamefont{D.~A.} \bibnamefont{Duncan}},
  \bibinfo{author}{\bibfnamefont{P.~K.} \bibnamefont{Thakur}},
  \bibinfo{author}{\bibfnamefont{T.-L.} \bibnamefont{Lee}},
  \bibinfo{author}{\bibfnamefont{A.~P.} \bibnamefont{Seitsonen}},
  \bibinfo{author}{\bibfnamefont{J.~V.} \bibnamefont{Barth}},
  \bibnamefont{et~al.}, \bibinfo{journal}{ACS Nano}
  \textbf{\bibinfo{volume}{11}}, \bibinfo{pages}{9151} (\bibinfo{year}{2017}).

\bibitem[{\citenamefont{Brülke et~al.}(2017)\citenamefont{Brülke,
  Heepenstrick, Humberg, Krieger, Sokolowski, Weiß, Tautz, and
  Soubatch}}]{Brulke2017}
\bibinfo{author}{\bibfnamefont{C.}~\bibnamefont{Brülke}},
  \bibinfo{author}{\bibfnamefont{T.}~\bibnamefont{Heepenstrick}},
  \bibinfo{author}{\bibfnamefont{N.}~\bibnamefont{Humberg}},
  \bibinfo{author}{\bibfnamefont{I.}~\bibnamefont{Krieger}},
  \bibinfo{author}{\bibfnamefont{M.}~\bibnamefont{Sokolowski}},
  \bibinfo{author}{\bibfnamefont{S.}~\bibnamefont{Weiß}},
  \bibinfo{author}{\bibfnamefont{F.~S.} \bibnamefont{Tautz}}, \bibnamefont{and}
  \bibinfo{author}{\bibfnamefont{S.}~\bibnamefont{Soubatch}},
  \bibinfo{journal}{J. Phys. Chem. C} \textbf{\bibinfo{volume}{121}},
  \bibinfo{pages}{23964} (\bibinfo{year}{2017}).

\bibitem[{\citenamefont{Bunk et~al.}(2007)\citenamefont{Bunk, Corso, Martoccia,
  Herger, Willmott, Patterson, Osterwalder, van~der Veen, and
  Greber}}]{Bunk2007}
\bibinfo{author}{\bibfnamefont{O.}~\bibnamefont{Bunk}},
  \bibinfo{author}{\bibfnamefont{M.}~\bibnamefont{Corso}},
  \bibinfo{author}{\bibfnamefont{D.}~\bibnamefont{Martoccia}},
  \bibinfo{author}{\bibfnamefont{R.}~\bibnamefont{Herger}},
  \bibinfo{author}{\bibfnamefont{P.}~\bibnamefont{Willmott}},
  \bibinfo{author}{\bibfnamefont{B.}~\bibnamefont{Patterson}},
  \bibinfo{author}{\bibfnamefont{J.}~\bibnamefont{Osterwalder}},
  \bibinfo{author}{\bibfnamefont{J.}~\bibnamefont{van~der Veen}},
  \bibnamefont{and} \bibinfo{author}{\bibfnamefont{T.}~\bibnamefont{Greber}},
  \bibinfo{journal}{Surf. Sci.} \textbf{\bibinfo{volume}{601}},
  \bibinfo{pages}{L7 } (\bibinfo{year}{2007}), ISSN \bibinfo{issn}{0039-6028},
  \urlprefix\url{http://www.sciencedirect.com/science/article/pii/S0039602806011551}.

\bibitem[{\citenamefont{Laskowski et~al.}(2007)\citenamefont{Laskowski, Blaha,
  Gallauner, and Schwarz}}]{Laskowski2007}
\bibinfo{author}{\bibfnamefont{R.}~\bibnamefont{Laskowski}},
  \bibinfo{author}{\bibfnamefont{P.}~\bibnamefont{Blaha}},
  \bibinfo{author}{\bibfnamefont{T.}~\bibnamefont{Gallauner}},
  \bibnamefont{and} \bibinfo{author}{\bibfnamefont{K.}~\bibnamefont{Schwarz}},
  \bibinfo{journal}{Phys. Rev. Lett.} \textbf{\bibinfo{volume}{98}},
  \bibinfo{pages}{106802} (\bibinfo{year}{2007}),
  \urlprefix\url{http://link.aps.org/doi/10.1103/PhysRevLett.98.106802}.

\bibitem[{\citenamefont{McKee et~al.}(2015)\citenamefont{McKee, Meunier, and
  Xu}}]{McKee2015}
\bibinfo{author}{\bibfnamefont{W.~C.} \bibnamefont{McKee}},
  \bibinfo{author}{\bibfnamefont{V.}~\bibnamefont{Meunier}}, \bibnamefont{and}
  \bibinfo{author}{\bibfnamefont{Y.}~\bibnamefont{Xu}}, \bibinfo{journal}{Surf.
  Sci.} \textbf{\bibinfo{volume}{642}}, \bibinfo{pages}{L16}
  (\bibinfo{year}{2015}).

\bibitem[{\citenamefont{Koch et~al.}(2012)\citenamefont{Koch, Langer, Kawai,
  Meyer, and Glatzel}}]{KochS12JPCM}
\bibinfo{author}{\bibfnamefont{S.}~\bibnamefont{Koch}},
  \bibinfo{author}{\bibfnamefont{M.}~\bibnamefont{Langer}},
  \bibinfo{author}{\bibfnamefont{S.}~\bibnamefont{Kawai}},
  \bibinfo{author}{\bibfnamefont{E.}~\bibnamefont{Meyer}}, \bibnamefont{and}
  \bibinfo{author}{\bibfnamefont{T.}~\bibnamefont{Glatzel}},
  \bibinfo{journal}{J. Phys.: Condens. Matter} \textbf{\bibinfo{volume}{24}},
  \bibinfo{pages}{314212} (\bibinfo{year}{2012}).

\bibitem[{\citenamefont{Dubout et~al.}(2016)\citenamefont{Dubout, Calleja,
  Sclauzero, Etzkorn, Lehnert, Claude, Papagno, Natterer, Patthey, Rusponi
  et~al.}}]{Dubout2016}
\bibinfo{author}{\bibfnamefont{Q.}~\bibnamefont{Dubout}},
  \bibinfo{author}{\bibfnamefont{F.}~\bibnamefont{Calleja}},
  \bibinfo{author}{\bibfnamefont{G.}~\bibnamefont{Sclauzero}},
  \bibinfo{author}{\bibfnamefont{M.}~\bibnamefont{Etzkorn}},
  \bibinfo{author}{\bibfnamefont{A.}~\bibnamefont{Lehnert}},
  \bibinfo{author}{\bibfnamefont{L.}~\bibnamefont{Claude}},
  \bibinfo{author}{\bibfnamefont{M.}~\bibnamefont{Papagno}},
  \bibinfo{author}{\bibfnamefont{F.~D.} \bibnamefont{Natterer}},
  \bibinfo{author}{\bibfnamefont{F.}~\bibnamefont{Patthey}},
  \bibinfo{author}{\bibfnamefont{S.}~\bibnamefont{Rusponi}},
  \bibnamefont{et~al.}, \bibinfo{journal}{New J. Phys.}
  \textbf{\bibinfo{volume}{18}}, \bibinfo{pages}{103027}
  (\bibinfo{year}{2016}),
  \urlprefix\url{http://stacks.iop.org/1367-2630/18/i=10/a=103027}.

\bibitem[{\citenamefont{Farwick~zum Hagen
  et~al.}(2016)\citenamefont{Farwick~zum Hagen, Zimmermann, Silva, Schlueter,
  Atodiresei, Jolie, Martínez-Galera, Dombrowski, Schröder, Will
  et~al.}}]{Hagen2016}
\bibinfo{author}{\bibfnamefont{F.~H.} \bibnamefont{Farwick~zum Hagen}},
  \bibinfo{author}{\bibfnamefont{D.~M.} \bibnamefont{Zimmermann}},
  \bibinfo{author}{\bibfnamefont{C.~C.} \bibnamefont{Silva}},
  \bibinfo{author}{\bibfnamefont{C.}~\bibnamefont{Schlueter}},
  \bibinfo{author}{\bibfnamefont{N.}~\bibnamefont{Atodiresei}},
  \bibinfo{author}{\bibfnamefont{W.}~\bibnamefont{Jolie}},
  \bibinfo{author}{\bibfnamefont{A.~J.} \bibnamefont{Martínez-Galera}},
  \bibinfo{author}{\bibfnamefont{D.}~\bibnamefont{Dombrowski}},
  \bibinfo{author}{\bibfnamefont{U.~A.} \bibnamefont{Schröder}},
  \bibinfo{author}{\bibfnamefont{M.}~\bibnamefont{Will}}, \bibnamefont{et~al.},
  \bibinfo{journal}{ACS Nano} \textbf{\bibinfo{volume}{10}},
  \bibinfo{pages}{11012} (\bibinfo{year}{2016}).

\bibitem[{\citenamefont{Ohtomo et~al.}(2017)\citenamefont{Ohtomo, Yamauchi,
  Sun, Kuzubov, Mikhaleva, Avramov, Entani, Matsumoto, Naramoto, and
  Sakai}}]{OhtomoM17Ns}
\bibinfo{author}{\bibfnamefont{M.}~\bibnamefont{Ohtomo}},
  \bibinfo{author}{\bibfnamefont{Y.}~\bibnamefont{Yamauchi}},
  \bibinfo{author}{\bibfnamefont{X.}~\bibnamefont{Sun}},
  \bibinfo{author}{\bibfnamefont{A.~A.} \bibnamefont{Kuzubov}},
  \bibinfo{author}{\bibfnamefont{N.~S.} \bibnamefont{Mikhaleva}},
  \bibinfo{author}{\bibfnamefont{P.~V.} \bibnamefont{Avramov}},
  \bibinfo{author}{\bibfnamefont{S.}~\bibnamefont{Entani}},
  \bibinfo{author}{\bibfnamefont{Y.}~\bibnamefont{Matsumoto}},
  \bibinfo{author}{\bibfnamefont{H.}~\bibnamefont{Naramoto}}, \bibnamefont{and}
  \bibinfo{author}{\bibfnamefont{S.}~\bibnamefont{Sakai}},
  \bibinfo{journal}{Nanoscale} \textbf{\bibinfo{volume}{9}},
  \bibinfo{pages}{2369} (\bibinfo{year}{2017}),
  \urlprefix\url{http://dx.doi.org/10.1039/C6NR06308J}.

\bibitem[{\citenamefont{Fadley}(1987)}]{Fadley1987}
\bibinfo{author}{\bibfnamefont{C.~S.} \bibnamefont{Fadley}},
  \bibinfo{journal}{Phys. Scr.} \textbf{\bibinfo{volume}{T17}},
  \bibinfo{pages}{39} (\bibinfo{year}{1987}).

\bibitem[{\citenamefont{Woodruff}(2007)}]{Woodruff2007}
\bibinfo{author}{\bibfnamefont{D.}~\bibnamefont{Woodruff}},
  \bibinfo{journal}{Surf. Sci. Rep.} \textbf{\bibinfo{volume}{62}},
  \bibinfo{pages}{1 } (\bibinfo{year}{2007}), ISSN \bibinfo{issn}{0167-5729},
  \urlprefix\url{http://www.sciencedirect.com/science/article/pii/S0167572906000902}.

\bibitem[{\citenamefont{Muntwiler et~al.}(2017)\citenamefont{Muntwiler, Zhang,
  Stania, Matsui, Oberta, Flechsig, Patthey, Quitmann, Glatzel, Widmer
  et~al.}}]{MuntwilerM17JSR}
\bibinfo{author}{\bibfnamefont{M.}~\bibnamefont{Muntwiler}},
  \bibinfo{author}{\bibfnamefont{J.}~\bibnamefont{Zhang}},
  \bibinfo{author}{\bibfnamefont{R.}~\bibnamefont{Stania}},
  \bibinfo{author}{\bibfnamefont{F.}~\bibnamefont{Matsui}},
  \bibinfo{author}{\bibfnamefont{P.}~\bibnamefont{Oberta}},
  \bibinfo{author}{\bibfnamefont{U.}~\bibnamefont{Flechsig}},
  \bibinfo{author}{\bibfnamefont{L.}~\bibnamefont{Patthey}},
  \bibinfo{author}{\bibfnamefont{C.}~\bibnamefont{Quitmann}},
  \bibinfo{author}{\bibfnamefont{T.}~\bibnamefont{Glatzel}},
  \bibinfo{author}{\bibfnamefont{R.}~\bibnamefont{Widmer}},
  \bibnamefont{et~al.}, \bibinfo{journal}{J. Synchrotron Rad.}
  \textbf{\bibinfo{volume}{24}}, \bibinfo{pages}{354} (\bibinfo{year}{2017}),
  \urlprefix\url{https://doi.org/10.1107/S1600577516018646}.

\bibitem[{\citenamefont{Hemmi et~al.}(2019)\citenamefont{Hemmi, Cun, Tocci,
  Epprecht, Stel, Lingenfelder, de~Lima, Muntwiler, Osterwalder, Iannuzzi
  et~al.}}]{HemmiA19NL}
\bibinfo{author}{\bibfnamefont{A.}~\bibnamefont{Hemmi}},
  \bibinfo{author}{\bibfnamefont{H.}~\bibnamefont{Cun}},
  \bibinfo{author}{\bibfnamefont{G.}~\bibnamefont{Tocci}},
  \bibinfo{author}{\bibfnamefont{A.}~\bibnamefont{Epprecht}},
  \bibinfo{author}{\bibfnamefont{B.}~\bibnamefont{Stel}},
  \bibinfo{author}{\bibfnamefont{M.}~\bibnamefont{Lingenfelder}},
  \bibinfo{author}{\bibfnamefont{L.~H.} \bibnamefont{de~Lima}},
  \bibinfo{author}{\bibfnamefont{M.}~\bibnamefont{Muntwiler}},
  \bibinfo{author}{\bibfnamefont{J.}~\bibnamefont{Osterwalder}},
  \bibinfo{author}{\bibfnamefont{M.}~\bibnamefont{Iannuzzi}},
  \bibnamefont{et~al.}, \bibinfo{journal}{Nano Lett.}
  \textbf{\bibinfo{volume}{19}}, \bibinfo{pages}{5998} (\bibinfo{year}{2019}).

\bibitem[{\citenamefont{Garc\'{\i}a~de Abajo
  et~al.}(2001)\citenamefont{Garc\'{\i}a~de Abajo, Van~Hove, and
  Fadley}}]{Abajo2001}
\bibinfo{author}{\bibfnamefont{F.~J.} \bibnamefont{Garc\'{\i}a~de Abajo}},
  \bibinfo{author}{\bibfnamefont{M.~A.} \bibnamefont{Van~Hove}},
  \bibnamefont{and} \bibinfo{author}{\bibfnamefont{C.~S.}
  \bibnamefont{Fadley}}, \bibinfo{journal}{Phys. Rev. B}
  \textbf{\bibinfo{volume}{63}}, \bibinfo{pages}{075404}
  (\bibinfo{year}{2001}),
  \urlprefix\url{http://link.aps.org/doi/10.1103/PhysRevB.63.075404}.

\bibitem[{\citenamefont{Hofmann et~al.}(1994)\citenamefont{Hofmann, Schindler,
  Bao, Bradshaw, and Woodruff}}]{Hofmann1994}
\bibinfo{author}{\bibfnamefont{P.}~\bibnamefont{Hofmann}},
  \bibinfo{author}{\bibfnamefont{K.-M.} \bibnamefont{Schindler}},
  \bibinfo{author}{\bibfnamefont{S.}~\bibnamefont{Bao}},
  \bibinfo{author}{\bibfnamefont{A.~M.} \bibnamefont{Bradshaw}},
  \bibnamefont{and} \bibinfo{author}{\bibfnamefont{D.~P.}
  \bibnamefont{Woodruff}}, \bibinfo{journal}{Nature}
  \textbf{\bibinfo{volume}{368}}, \bibinfo{pages}{131} (\bibinfo{year}{1994}),
  \urlprefix\url{https://doi.org/10.1038/368131a0}.

\bibitem[{\citenamefont{Duncan et~al.}(2012)\citenamefont{Duncan, Choi, and
  Woodruff}}]{Duncan2012}
\bibinfo{author}{\bibfnamefont{D.}~\bibnamefont{Duncan}},
  \bibinfo{author}{\bibfnamefont{J.}~\bibnamefont{Choi}}, \bibnamefont{and}
  \bibinfo{author}{\bibfnamefont{D.}~\bibnamefont{Woodruff}},
  \bibinfo{journal}{Surf. Sci.} \textbf{\bibinfo{volume}{606}},
  \bibinfo{pages}{278 } (\bibinfo{year}{2012}), ISSN \bibinfo{issn}{0039-6028},
  \urlprefix\url{//www.sciencedirect.com/science/article/pii/S0039602811003979}.

\bibitem[{\citenamefont{Usachov et~al.}(2018)\citenamefont{Usachov, Tarasov,
  Bokai, Shevelev, Vilkov, Petukhov, Rybkin, Ogorodnikov, Kuznetsov, Muntwiler
  et~al.}}]{Usachov2018}
\bibinfo{author}{\bibfnamefont{D.~Y.} \bibnamefont{Usachov}},
  \bibinfo{author}{\bibfnamefont{A.~V.} \bibnamefont{Tarasov}},
  \bibinfo{author}{\bibfnamefont{K.~A.} \bibnamefont{Bokai}},
  \bibinfo{author}{\bibfnamefont{V.~O.} \bibnamefont{Shevelev}},
  \bibinfo{author}{\bibfnamefont{O.~Y.} \bibnamefont{Vilkov}},
  \bibinfo{author}{\bibfnamefont{A.~E.} \bibnamefont{Petukhov}},
  \bibinfo{author}{\bibfnamefont{A.~G.} \bibnamefont{Rybkin}},
  \bibinfo{author}{\bibfnamefont{I.~I.} \bibnamefont{Ogorodnikov}},
  \bibinfo{author}{\bibfnamefont{M.~V.} \bibnamefont{Kuznetsov}},
  \bibinfo{author}{\bibfnamefont{M.}~\bibnamefont{Muntwiler}},
  \bibnamefont{et~al.}, \bibinfo{journal}{Phys. Rev. B}
  \textbf{\bibinfo{volume}{98}}, \bibinfo{pages}{195438}
  (\bibinfo{year}{2018}),
  \urlprefix\url{https://link.aps.org/doi/10.1103/PhysRevB.98.195438}.

\bibitem[{\citenamefont{de~Lima et~al.}(2013)\citenamefont{de~Lima, de~Siervo,
  Landers, Viana, Goncalves, Lacerda, and H\"aberle}}]{deLima2013}
\bibinfo{author}{\bibfnamefont{L.~H.} \bibnamefont{de~Lima}},
  \bibinfo{author}{\bibfnamefont{A.}~\bibnamefont{de~Siervo}},
  \bibinfo{author}{\bibfnamefont{R.}~\bibnamefont{Landers}},
  \bibinfo{author}{\bibfnamefont{G.~A.} \bibnamefont{Viana}},
  \bibinfo{author}{\bibfnamefont{A.~M.~B.} \bibnamefont{Goncalves}},
  \bibinfo{author}{\bibfnamefont{R.~G.} \bibnamefont{Lacerda}},
  \bibnamefont{and}
  \bibinfo{author}{\bibfnamefont{P.}~\bibnamefont{H\"aberle}},
  \bibinfo{journal}{Phys. Rev. B} \textbf{\bibinfo{volume}{87}},
  \bibinfo{pages}{081403} (\bibinfo{year}{2013}),
  \urlprefix\url{https://link.aps.org/doi/10.1103/PhysRevB.87.081403}.

\bibitem[{\citenamefont{de~Lima et~al.}(2014)\citenamefont{de~Lima, Handschak,
  Schonbohm, Landers, Westphal, and de~Siervo}}]{deLima2014}
\bibinfo{author}{\bibfnamefont{L.~H.} \bibnamefont{de~Lima}},
  \bibinfo{author}{\bibfnamefont{D.}~\bibnamefont{Handschak}},
  \bibinfo{author}{\bibfnamefont{F.}~\bibnamefont{Schonbohm}},
  \bibinfo{author}{\bibfnamefont{R.}~\bibnamefont{Landers}},
  \bibinfo{author}{\bibfnamefont{C.}~\bibnamefont{Westphal}}, \bibnamefont{and}
  \bibinfo{author}{\bibfnamefont{A.}~\bibnamefont{de~Siervo}},
  \bibinfo{journal}{Chem. Commun.} \textbf{\bibinfo{volume}{50}},
  \bibinfo{pages}{13571} (\bibinfo{year}{2014}),
  \urlprefix\url{http://dx.doi.org/10.1039/C4CC05005C}.

\bibitem[{\citenamefont{de~Lima et~al.}(2016)\citenamefont{de~Lima, Barreto,
  Landers, and de~Siervo}}]{deLima2016}
\bibinfo{author}{\bibfnamefont{L.~H.} \bibnamefont{de~Lima}},
  \bibinfo{author}{\bibfnamefont{L.}~\bibnamefont{Barreto}},
  \bibinfo{author}{\bibfnamefont{R.}~\bibnamefont{Landers}}, \bibnamefont{and}
  \bibinfo{author}{\bibfnamefont{A.}~\bibnamefont{de~Siervo}},
  \bibinfo{journal}{Phys. Rev. B} \textbf{\bibinfo{volume}{93}},
  \bibinfo{pages}{035448} (\bibinfo{year}{2016}),
  \urlprefix\url{https://link.aps.org/doi/10.1103/PhysRevB.93.035448}.

\bibitem[{\citenamefont{Sclauzero and Pasquarello}(2012)}]{Sclauzero2012}
\bibinfo{author}{\bibfnamefont{G.}~\bibnamefont{Sclauzero}} \bibnamefont{and}
  \bibinfo{author}{\bibfnamefont{A.}~\bibnamefont{Pasquarello}},
  \bibinfo{journal}{Phys. Rev. B} \textbf{\bibinfo{volume}{85}},
  \bibinfo{pages}{161405} (\bibinfo{year}{2012}),
  \urlprefix\url{https://link.aps.org/doi/10.1103/PhysRevB.85.161405}.

\end{thebibliography}

\end{document}